\documentclass[aps,prl,twocolumn,superscriptaddress,showpacs,floatfix]{revtex4-2}
\usepackage{amsmath,amssymb,amsfonts,float,graphics,epsfig,epstopdf,color,verbatim,tabularx,bm,multirow}
\usepackage{hyperref}
\usepackage[cmyk]{xcolor}
\usepackage{bbold}
\hypersetup{
colorlinks = true,
linkcolor = [rgb]{0.70,0.13,0.13},
citecolor = [rgb]{0.13,0.55,0.13},
urlcolor = [rgb]{0.25,0.41,0.88}}

\DeclareSymbolFont{sfletters}{OML}{cmbrm}{m}{it}
\DeclareMathSymbol{\sfeps}{\mathord}{sfletters}{"22}

\graphicspath{{Figures/}}

\begin{document}

\title{Universal collective Larmor-Silin mode emerging in magnetized correlated Dirac fermions}

\author{Chuang Chen}
\affiliation{State Key Laboratory of Surface Physics, Fudan University, Shanghai 200433, China}
\affiliation{Center for Field Theory and Particle Physics, Department of Physics, Fudan University, Shanghai 200433, China}

\author{Yuan Da Liao}
\affiliation{Department of Physics and HK Institute of Quantum Science \& Technology, The University of Hong Kong, Pokfulam Road, Hong Kong SAR, China}

\author{Chengkang Zhou}
\affiliation{Department of Physics and HK Institute of Quantum Science \& Technology, The University of Hong Kong, Pokfulam Road, Hong Kong SAR, China}

\author{Gaopei Pan}
\affiliation{Department of Physics and HK Institute of Quantum Science \& Technology, The University of Hong Kong, Pokfulam Road, Hong Kong SAR, China}
\affiliation{Institut f\"ur Theoretische Physik und Astrophysik and W\"urzburg-Dresden Cluster of Excellence ct.qmat, Universit\"at W\"urzburg, 97074 W\"urzburg, Germany}

\author{Zi Yang Meng}
\email{zymeng@hku.hk}
\affiliation{Department of Physics and HK Institute of Quantum Science \& Technology, The University of Hong Kong, Pokfulam Road, Hong Kong SAR, China}

\author{Yang Qi}
\email{qiyang@fudan.edu.cn}
\affiliation{State Key Laboratory of Surface Physics, Fudan University, Shanghai 200433, China}
\affiliation{Center for Field Theory and Particle Physics, Department of Physics, Fudan University, Shanghai 200433, China}
\affiliation{Collaborative Innovation Center of Advanced Microstructures, Nanjing 210093, China}

\date{\today}

\begin{abstract}
Employing large-scale quantum Monte Carlo simulations, we find that in the magnetized interacting Dirac fermion model there emerges a universal collective Larmor-Silin spin wave mode in the transverse dynamical spin susceptibility. Such mode purely originates from the interaction among Dirac fermions and distinguishes itself from the usual particle-hole continuum with finite lifetime and clear dispersion, both at small and large momenta in a large portion of the Brillouin zone. Our unbiased numerical results offer the dynamic signature of this collective excitation in interacting Dirac fermion systems, and provide experimental guidance for inelastic neutron scattering, electron spin resonance, and other spectroscopic approaches in the investigation of such universal collective modes in quantum Moir\'e materials, topological insulators, and quantum spin liquid materials under magnetic field, with quintessential interaction nature beyond the commonly assumed noninteracting Dirac fermion or spinon approximations.
\end{abstract}

\maketitle

\noindent{\textcolor{blue}{\it Introduction.}---} Collective excitations offer
the fingerprint of quantum many-body systems, e.g., the spin wave in the magnetically ordered systems~\cite{nearlyShao2017, cftd2001,
  plihal1999spinwave, vollmer2003spinwave, prabhakar2009spin}, the roton mode in the
superfluid~\cite{feynmanSuperfluidity1957,liKosterlitz2020,landau2018theory,
  bisset2013fingerprinting, blakie2012roton, rybalko2010microwave}, excitons and magnetorotons in the quantum moir\'e
materials, integer and fractional quantum (anomalous) Hall systems~\cite{girvinMagneto1986,linExciton2022,panThermodynamic2023,luThermodynamic2023,lu_2024,luInteraction2024,nguyenMultiple2022,kumarNeutral2022} and many others. And it is oftentimes the case that the identification of new collective excitations
provides the decisive understanding of the physical nature of the corresponding quantum many-body ground states.
%the new understanding of the primary degrees of freedom that govern the
%behavior of the physical systems and reveal the underlying mechanism at working.

The situation becomes subtle in highly entangled quantum matter, where
the collective excitations are harder to identify in an unbiased manner. For
example, in the study of quantum spin liquid (QSL) states, it is well known that
the identification of the unique aspect of a QSL state from a specific
experimental signature is difficult, as there usually exists multiple
explanations for the same experimental
data~\cite{BalentSpinliquids2010,SavaryDisorder2017,ZhouYQUantum2017,BalentSpinliquids2010}.
One crucial property of QSL is the emergence of fractionalized excitations, such
as spinons (carrying a spin of 1/2 but charge neutral), which are elementary
quasiparticles carrying topological characteristics and interacting with an
emergent gauge
field~\cite{KivelsonSATopological1987,WenXGMean-field1991,WenXGZoo2017,GYSunDynamical2018,wangFractionalized2021,xuMonte2019,wangDynamics2019,huangDynamics2018,wangVestigial2021}.
In Dirac QSLs, spinons can exhibit a conical dispersion~\cite{zengDirac2023},
resembling Dirac cones in the electronic band structures of graphene,
topological insulators, and many two-dimensional (2D) materials~\cite{GeimAKTROG2007,HasanMZTI2010,VafekDirac2014,makSemiconductor2022}. %As a result, the spinons in Dirac QSLs represent a novel type of Dirac quasiparticles without the charge degree of freedom. 

\begin{figure}[htp!]
	\includegraphics[width=\columnwidth]{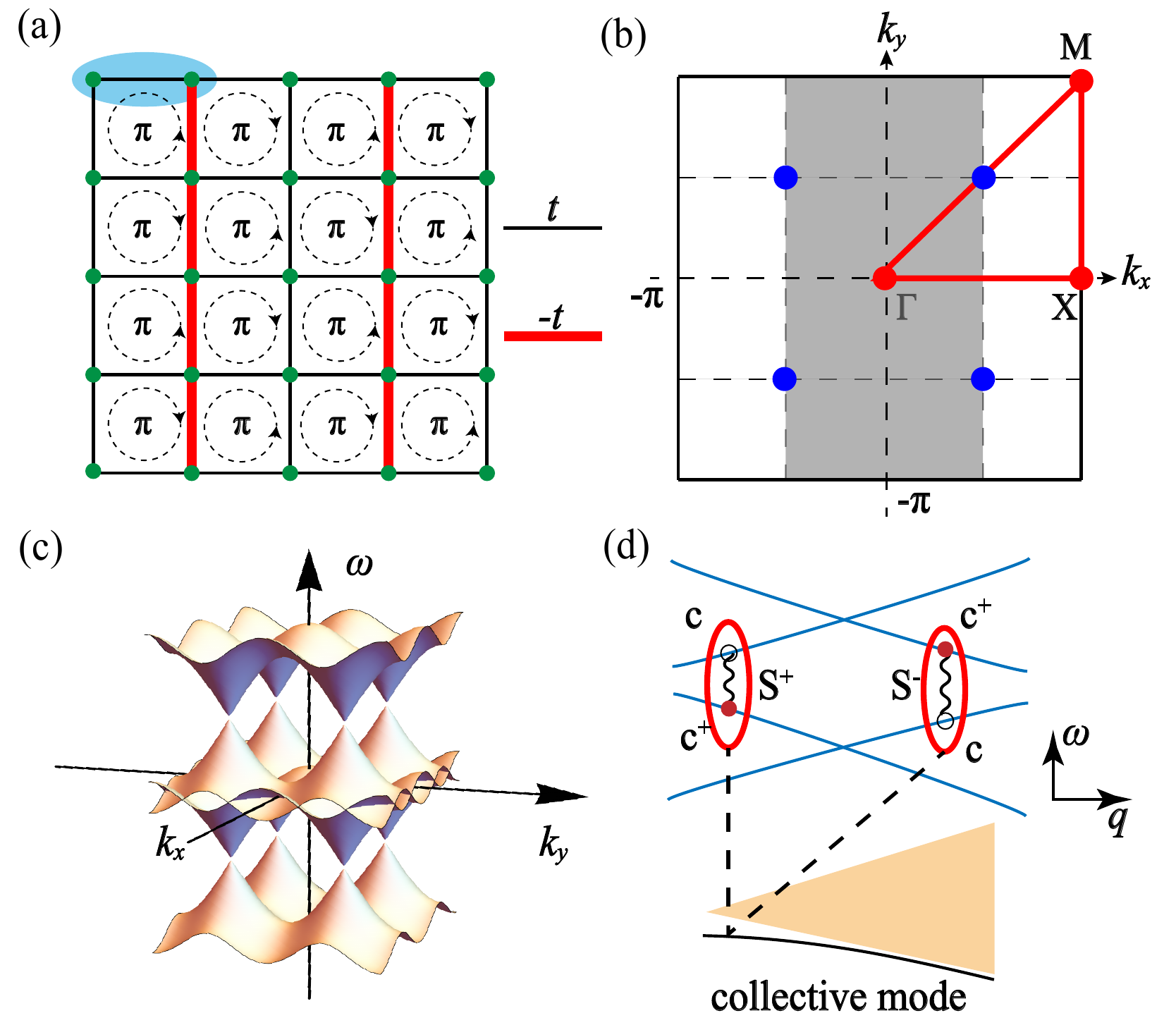}
	\caption{(a) Schematic plot of the Hamiltonian Eq.~\eqref{eq:Ham}, the
    $\pi$-flux model on the square lattice. The black lines stand for then
    hopping sign of the fermion with $t_{ij} = +1$, while the red lines are
    $t_{ij} = -1$. Such arrangement ensures that the flux threading each
    plaquette is $\pi$, giving Dirac cones at zero magnetic field. The blue
    ellipse enclosing two lattice sites is the unit cell. (b) BZ of the original
    square lattice (white) and the folded BZ (gray) with blue ellipse as unit
    cell. The path denoted by the red line connects high symmetry momenta
    $\Gamma(0,0)$, $X(\pi, 0)$ and $M(\pi, \pi)$ in BZ for the original square
    lattice. (c) The split Fermi surface of the $\pi$-flux model under external magnetic field in $z$ direction. (d) Schematic plot that depicts the origin of the collective mode. The coherent excitation of spin flip $S^+$ and $S^-$ in the model gives rise to the collective mode.}
	\label{fig:fig1}
\end{figure}

While detecting single spinon excitations is challenging, their collective
excitations -- two spinon excitations with a total spin quantum number $S$ = 1  can lead to a spin continuum spectrum that can be detected through inelastic neutron scattering techniques. Such characteristic collective continuum spectra have been reported in materials such as the kagome lattice antiferromagnets  ZnCu$_3$(OH)$_6$Cl$_2$~\cite{hanFractionalized2012}, Cu$_3$Zn(OH)$_6$FBr~\cite{weiEvidence2017} and more recently YCu$_3$(OH)$_6$Br$_2$[Br$_{1-x}$(OH)${_x}$]~\cite{zengPossible2022,zengDirac2023,zheng2023unconventional,georgopoulouMagnetically2023,liuGapless2022}.

However, in these studies of QSL, one often assumes the fractionalized
spinons are nearly free particles and computes their collective modes (continuum
spectrum) under such assumption, i.e., the convolution of two independent
spectra of a single spinon~\cite{arikawa2006spinon,ran2009spontaneous,
shen_evidence_2016}. But in reality, it is obvious that the spinons experience strong interactions, mediated by the fluctuations of gauge fields~\cite{DHKim1997,YRan2007,xuMonte2019,wangDynamics2019,yanTopological2021,maDynamical2018,wangFractionalized2021}, and the free spinon assumption is oversimplified and often leads to contradictions or controversies, when trying to interpret the experimental data and make predictions~\cite{maSpin2018,zhuTopography2018}.

Therefore, one needs to either solve the interacting problem completely, usually
via unbiased numerical approaches such as quantum Monte Carlo
(QMC)~\cite{wangFractionalized2021,xuMonte2019,wangDynamics2019,yanTopological2021,huangDynamics2018,GYSunDynamical2018,wangVestigial2021},
or find new signatures which are robust beyond the nearly free approximations, even when the interaction effect is included. But neither of these {\it two strategies} is easy and the progresses are usually made in a case by case manner. Recently, an interesting proposal of the latter strategy came to our attention. In Refs.~\cite{keselmanDynamical2020,balentsCollective2020,agarwalCollective2023}, the authors propose a new collective "spinon spin wave" mode to emerge in the transverse dynamical spin susceptibility -- different from the usual spin continuum spectra -- in the 2D Dirac QSL subjected to an applied
Zeeman field, and tested their proposal rigorously in a one-dimensional
Heisenberg chain via density matrix renormalization group simulations and in 2D magnetized graphene in perturbative analysis.

The suggested collective modes date back to early investigations of the spin response of paramagnetic metals subjected to the external magnetic field, denoted as the transverse Larmor-Silin spin wave~\cite{silinOscillations1958,silinOscillations1959}, in the form of collective spin oscillations in itinerant interacting fermion systems~\cite{platzmanSpin1967,leggettSpin1970,vanloonLarmor2023,schultzObservation1967}. Phenomenologically, such Larmor-Silin spin wave is the transverse collective spin mode below the particle-hole continuum of, say, a single-band conductor with parabolic electron dispersion. This downward dispersing collective mode is the precession of the total magnetic moment originating from the Zeeman frequency at zero wave vector (as schematically shown in Fig.~\ref{fig:fig1} (d)) and of purely interacting nature. The spinon spin wave in Refs.~\cite{keselmanDynamical2020,balentsCollective2020,agarwalCollective2023}, can therefore be viewed as the modern version of the Larmor-Silin spin wave, in the context of magnetized graphene and 2D Dirac QSL.

\begin{figure}[htp!]
	\includegraphics[width=\columnwidth]{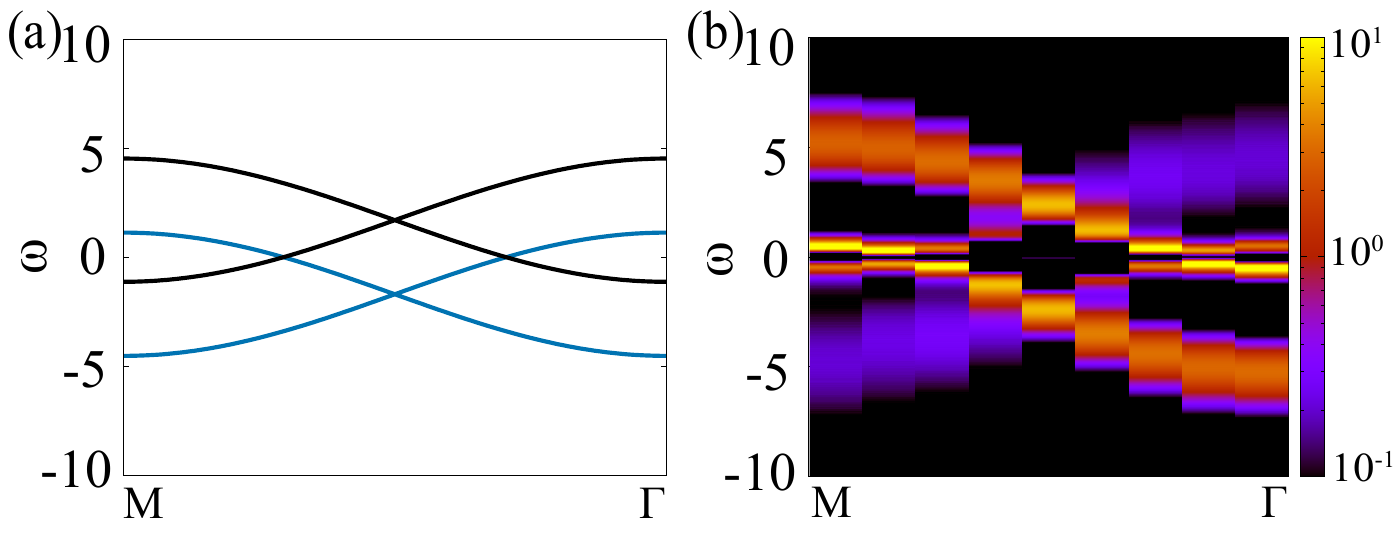}
	\caption{(a) Free fermion dispersion under magnetic field $B=3.4$. Because of
		polarization, the spin degeneracy is lifted and four bands are visible. (b)
		Single particle spectra $A(k, \omega)$ from QMC+SAC with
		parameter $U=2, B=3.4$ on a $16\times 16$ lattice. The gap at $(\frac{\pi}{2}, \frac{\pi}{2})$ is
		around 2.34, corresponding to an effective magnetic field
		$B_{\text{eff}}=4.68$ in Eq.~\eqref{eq:beff}.}
	\label{fig:fig2}
\end{figure}

Until now, the unbiased 2D lattice model verification of the spinon spin wave,
i.e., to carry out both the strategy of solving the interacting problem
completely via unbiased numerical calculation and the strategy of identifying a
new signature of the purely interaction-generated collective mode beyond the
nearly free approximation, has not been achieved. This is due, of course, to
both the lack of proper lattice model design and the numerical difficulties in
solving the interacting problem accurately. In this Letter, we finally achieve
both goals successfully, by employing large-scale QMC simulations, complemented
with random phase approximation (RPA) analysis, to find that in a concrete 2D
magnetized interacting Dirac fermion lattice model there emerges the universal
collective Larmor-Silin spin wave mode in the transverse dynamical spin
susceptibility. Beyond the perturbative
proposals~\cite{balentsCollective2020,agarwalCollective2023}, we find in the 2D
lattice model that the Larmor-Silin spin wave not only splits off from the usual "two-spinon continuum" at small momenta, but also universally appears inside and at the bottom of the continuum at large momenta, with richer renormalization effects and magnetic field dependence. Our unbiased numerical results offer the dynamic signature of this collective mode in interacting Dirac fermion systems, and might provide experimental relevance for its detection via spectroscopic measurements (for example inelastic neutron scattering) in quantum Moir\'e materials, topological insulators, and Dirac spin liquid materials, under magnetic field.

\begin{figure*}[htp!]
\begin{center}
\includegraphics[width=\textwidth]{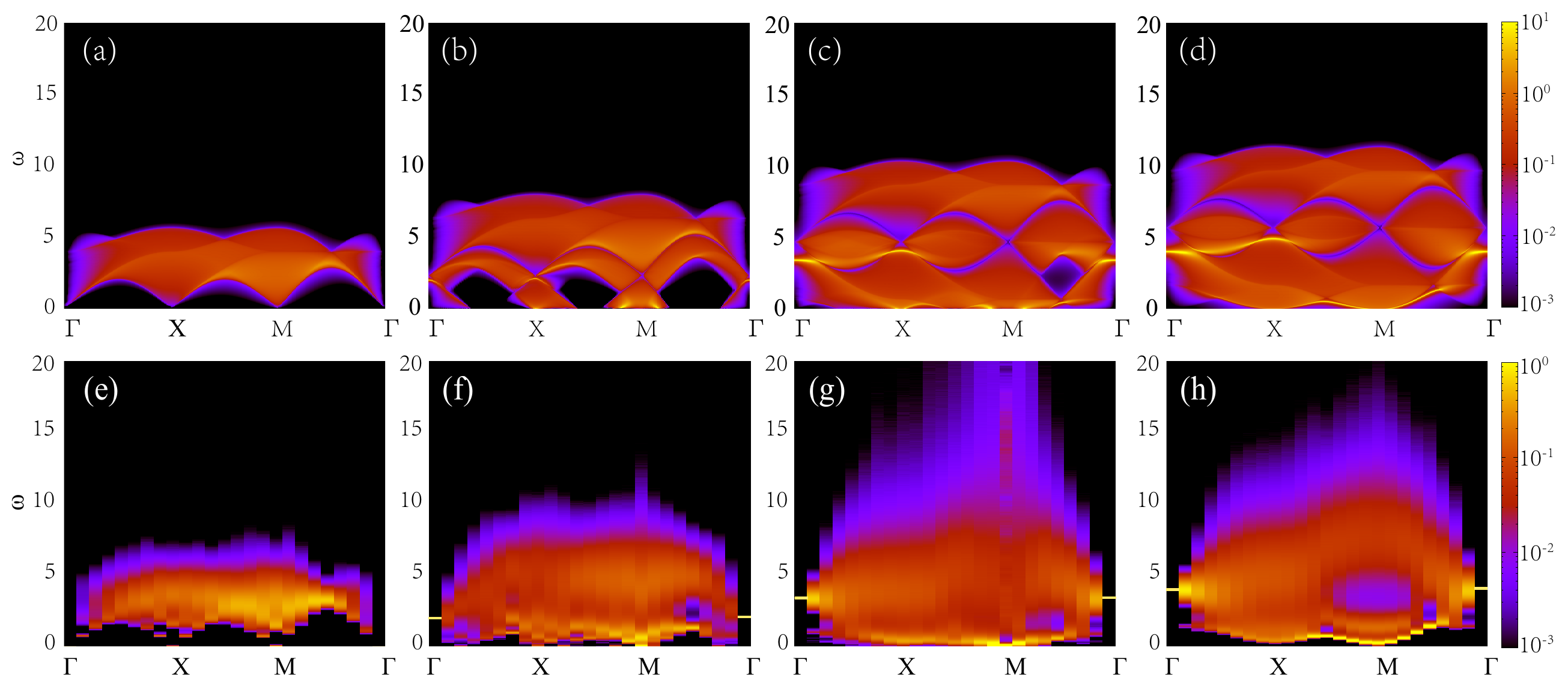}
\caption{ Transverse magnetic spectra $S_{\pm}(\mathbf{q}, \omega)$ under $z$ direction magnetic field at $U=2$.  Panels (a), (b), (c) and (d) are the spectra from RPA spin susceptibilities $\chi_{\pm}(\mathbf{q}, \omega)$ with $B=0, 2, 3.4,4$. Panels (e), (f), (g) and (h) are the spectra from QMC+SAC results with $B=0, 2,3.4, 4$. As magnetic field $B$ increases, the universal collective transverse mode emerges, not only close to $\Gamma$ but also at higher momenta and inside the spin continuum. The RPA and QMC spin spectra are consistent with each other.}
		\label{fig:fig3}
	\end{center}
\end{figure*}

\noindent{\textcolor{blue}{\it Model and Method.}---}
Our model has the following Hamiltonian on a 2D square lattice:
\begin{equation}
\begin{aligned}
  H &= H_0+H_U+H_B, \\
  H_0 &= -t \sum_{\langle i,j \rangle, \sigma} (t_{ij}c^{\dagger}_{i,\sigma} c_{j,\sigma} + h.c.)  \\
  H_U &=  \frac{U}{2} \sum_{i} (n_{i\uparrow} + n_{i\downarrow} - 1)^2 \\
  H_B & = -B \sum_{i}\frac{1}{2}(c^{\dagger}_{i,\uparrow}c_{i,\uparrow}  - c^{\dagger}_{i,\downarrow}c_{i,\downarrow})
\end{aligned}
\label{eq:Ham}
\end{equation}
and as shown in Fig.~\ref{fig:fig1} (a), the hopping sign $t_{ij}$ on black (red) bonds is $+1$ $(-1)$, ensuring the flux threading each plaquette
is equal to $\pi$. $\sigma=\uparrow, \downarrow$ denotes spin flavors. The blue ellipse denotes the unit cell, which contains two sites. The Brillouin zone (BZ), shown in Fig.~\ref{fig:fig1}(b),  is folded with $k_x \in [-\pi/2,
\pi/2)$ and $k_y \in [-\pi, \pi)$. The free dispersion has two independent Dirac points denoted by the blue solid points. Near the Dirac points, the dispersion has good linear form. The interaction in the model
is the on-site Hubbard repulsion with strength $U$. The
$S^z_i=\frac{1}{2}(c^{\dagger}_{i,\uparrow}c_{i,\uparrow}-c^{\dagger}_{i,\downarrow}c_{i,\downarrow})$
of the fermion is
coupled to an external magnetic field in $z$ direction with strength $B$. The
effect of magnetic field $B$ can be interpreted as that the spin up and spin down
fermion have chemical potential with opposite sign. 

Without external magnetic field, the $\pi$-flux Hubbard model with spin $1/2$
fermion has a Dirac semi-metal to antiferromagnetic N\'eel state quantum phase transition at finite $U_c$, with the latter state spontaneously breaking $SU(2)$ spin rotational
symmetry~\cite{bercx2009,OtsukaPiflux2002,ChangPiflux2012,ParisenPiflux2015,OtsukaPiflux2016}.
In the presence of Zeeman field $B$, there will also be a paramagnetic to
in-plane antiferromagnetic state transition at $U_c(B)$~\cite{bercx2009}.
However, since we are interested in the Larmor-Silin mode as a collective
transverse spin excitation without spontaneous magnetic order, our main results
are in the regime of $U<U_c(B)$ such that the spontaneous in-plane
antiferromagnetic state has not been established but the magnetized Dirac
fermions are nevertheless interacting. The magnetic spectra in the situation of
$U>U_c(B)$ are presented in the Supplemental Material (SM)~\cite{suppl} (see also references ~\cite{Olav2008,shao2023progress} therein).
With plaquette interactions, extended beyond on-site, it is also found via
large-scale QMC simulation that the model can host a $U(1)$ Dirac QSL phase, via the
nontrivial deconfined quantum critical
point~\cite{ouyangProjection2021,liaoDiracI2022,liaoDiracII2022,liaoDiracIII2022}.
It is fair to assert that the $\pi$-flux Dirac fermion model can be used to describe the spinon dispersion of a $U(1)$ Dirac QSL~\cite{Wenbook}. %In order to answer what might the spin spectra looks like for an $U(1)$ quantum spin liquid under external magnetic field, we exploit the model Eq.~\eqref{eq:Ham} to approximately represent an $U(1)$ QSL. 
Therefore, the spin spectra of our model can help to understand the spectrum of the $U(1)$
Dirac QSL under magnetic field. Admittedly, the present model does not have the
dynamical $U(1)$ gauge field and we shall leave it for future work.
%\cc{We would like to confirm the existence of Larmor-Silin spin wave in magnetized Dirac fermion model through our investigation. The spinon spin wave is the Larmor-Silin spin wave in model that couples the fermion with dynamical $U(1)$ gauge field.}

\noindent{\textcolor{blue}{\it RPA Analysis.}---}
First we investigate the model Eq.~\eqref{eq:Ham} with RPA following the work~\cite{balentsCollective2020}. The
detailed calculation scheme is given in Sec. I of SM~\cite{suppl}. %After Fourier transformation, we can obtain momentum space Hamiltonian $H_k$. 
We define magnetization as
$M=\frac{1}{2}(\frac{N_{\uparrow}-N_{\downarrow}}{N_{\uparrow}+N_{\downarrow}})$,
where $N_{\uparrow}$ ($N_{\downarrow}$) is the particle number of the up (down)
fermion. The system is at half filling,
$N_{\text{tot}} = N_{\uparrow} + N_{\downarrow} = L^2/2$, where $L$ is the linear system size. The contribution to the Green's function from Hubbard
interaction $U$ is approximately treated as Hartree energy shift. The fermion dispersions are shifted
by $-UM$ for up spin and $UM$ for down spin. In practice, we find it more appropriate to use renormalized interaction strength $U_{\text{eff}}$
in RPA calculation. As shown in Fig.~\ref{fig:fig2}, in QMC we can obtain the effective
magnetic field $B_\text{eff}$ which is two times the gap at the Dirac point $(\pi/2,
\pi/2)$ from single particle spectra $A(\mathbf{k}, \omega)$ . $B_\text{eff}$ is then related to $U_{\text{eff}}$
by
\begin{equation}
B_{\text{eff}} = B + U_{\text{eff}}M
\label{eq:beff}.
\end{equation}

The renormalized interaction strength $U_{\text{eff}}$ is then calculated as
$U_{\text{eff}}=\frac{B_{\text{eff}}-B}{M}$. Consider the effective Hamiltonian,
$\epsilon_{a,\sigma}(k)= (-1)^a 2t \sqrt{\cos^2 \mathbf{k}_x + \cos^2 \mathbf{k}_y}
- \alpha(\sigma) B_{\text{eff}}/2$, where  $a\in [1,2]$ denotes the band index and
$\alpha=+1(-1)$ for $\uparrow(\downarrow)$ spin . We then calculate the bare
transverse spin susceptibility $\chi^0_{\pm}$ with standard fermion loop
diagram. The RPA spin susceptibility $\chi_{\pm}$ is obtained by resummation of particle-hole ladder diagrams with $\chi^{0}_{\pm}$ as bare susceptibility~\cite{aronov1977spin}
\begin{equation}
  \chi_{\pm}(\mathbf{q}, i \omega_n) = \frac{\chi^{0}_{\pm}(\mathbf{q}, i \omega_n)}{1+U_{\text{eff}}\chi^{0}_{\pm}(\mathbf{q}, i \omega_n)}
\label{eq:chi}.
\end{equation}
After analytic continuation $i\omega_n \rightarrow \omega + i0^{+}$, the spin
spectra $S_{\pm}(\mathbf{q}, \omega)= -\text{Im} \chi_{\pm}(\mathbf{q},
\omega)$, which will be directly compared with QMC results.
 The existence of the universal collective mode is already manifested at RPA spin
susceptibility~\cite{aronov1977spin, balentsCollective2020} and the condition
for its emergence is discussed in detail in SM~\cite{suppl}.

\begin{figure}[htp!]
	\includegraphics[width=\columnwidth]{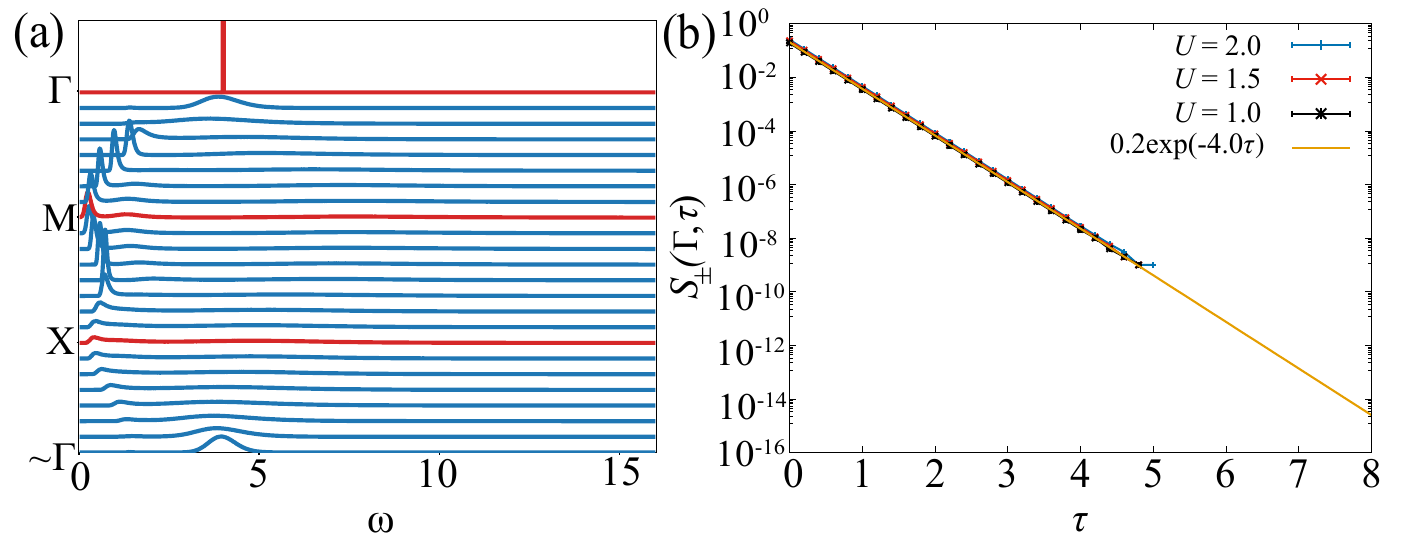}
	\caption{(a) Spectra $S_{\pm}(\mathbf{q},
		\omega)$ from QMC+SAC at $U=2, B=4$. The red lines highlight the high
		symmetry momentum points in BZ. (b) Logscale plot of imaginary-time spin correlation function $S_\pm(\mathbf{\Gamma}, \tau)$ from
		QMC simulation with $B=4$ and varying $U$.}
	\label{fig:fig4}
\end{figure}

\noindent{\textcolor{blue}{\it QMC Simulation and Results.}---}
We further investigate the model with finite-temperature determinant quantum
Monte Carlo (DQMC) simulations~\cite{bss1981,xuRevealing2019}. The model in Eq.~\eqref{eq:Ham} is particle-hole
symmetric and thus sign-problem free~\cite{wuSufficient2005,panSign2024}. We use the Hubbard-Stratonovich decomposition to decouple the interaction into
density channel and perform the simulations on a $16 \times 16$ square lattice
(at the inverse temperature $\beta t=L$ with $t=1$ as the energy unit) at $U=2$
with varying external magnetic field $B$. The external magnetic field $B$ enters
the DQMC as the chemical potential term for canonical ensemble simulations, with
opposite sign for opposite spin flavor, and still renders the simulation
sign-problem free. As discussed above,  at large $U>U_c(B)$, the system will break the $U(1)$
spin rotational symmetry and enter an antiferromagnetic XY (AFXY) ordered insulator phase~\cite{bercx2009}, but
we purposely stay in the paramagnetic phase to investigate the transverse
Larmor-Silin spin wave mode. The details of DQMC implementation and the magnetic spectra at $U>U_c(B)$ are given in Secs. II and III of SM~\cite{suppl}, respectively.

First we calculate the imaginary time Green's function $G_\sigma(\mathbf{k},\tau)$ and use the stochastic analytic continuation
(SAC)~\cite{sandvik1998sac,sandvik2016constrained,beach2004identifying,nearlyShao2017,GYSunDynamical2018,wangFractionalized2021,wangDynamics2019,maDynamical2018} to
obtain the real frequency single particle spectral function $A(\mathbf{k},
\omega)$. The result of
$A(\mathbf{k}, \omega)$ is shown in Fig.~\ref{fig:fig2} (b) comparing
with dispersion of the free fermion under the same external magnetic field in Fig.~\ref{fig:fig2} (a). As shown
in the Figure, the degeneracy of energy bands is lifted by the
external magnetic field $B$, giving rise to the pocket like Fermi surface shown in
Fig.~\ref{fig:fig1} (c). In our lattice simulation with $L=16, \beta=16, U=2,
B=3.4$, the QMC single particle spectrum shows that the magnetized system is
still in the metallic state, and the amount of energy shift is
larger compared to the free system with $U=0, B=3.4$, which can be attributed to
the renormalization effect of the Hubbard interaction. This further confirms the usage of effective Hubbard $U_{\text{eff}}$ as Hartree energy shift in RPA calculation.

Next we focus on the spin spectra of the magnetized Dirac fermions. We compute the dynamical transverse spin
correlation function
\begin{equation}
  S_{\pm}(\mathbf{q}, \tau) = \frac{1}{L^4} \sum_{i,j} \left< S^+_i (\tau) S^-_j(0) \right> e^{-i (r_i - r_j) \cdot \mathbf{q}}
\label{eq:Scr}.
\end{equation}
where $S^+_i = c^\dagger_{i,\uparrow} c_{i,\downarrow}$ and $S^-_i =
c^\dagger_{i,\downarrow} c_{i,\uparrow}$. The same as $A(\mathbf{k},\omega)$, we can obtain the
spin spectra $S_\pm(\mathbf{q}, \omega)$ from $S_\pm(\mathbf{q}, \tau)$
utilizing SAC, whose details are given in Sec. II of SM~\cite{suppl}.

The results are depicted in Fig.~\ref{fig:fig3}. The upper panels are RPA spin
spectra $S_{\pm}(\mathbf{q}, \omega)$, while the lower panels are QMC results,
with the parameters $U=2$, $L=\beta=16$. The RPA calculations are at zero
temperature with $50 \times 100$ momentum grid; considering the unit cell, the
BZ is folded with $k_x \in [-\pi/2,\pi/2)$ and $k_y \in [-\pi, \pi)$. The
effective $U_\text{eff}$ in the RPA panels are determined according to
Eq.~\eqref{eq:beff} with the help of single particle spectra $A(\mathbf{k},
\omega)$. For $B=2, 3.4, 4$, effective $U_\text{eff}=2.86, 1.69, 1.67$
respectively. Without magnetic field at $U=2$, the magnetization $\left< M
\right>=0$ in QMC, thus use $U_\text{eff}=U=2$ in RPA.
%\zym{(Here it is not clear how do you determine the $U_{eff}$ for
%panels (a), (b), (c), (d) in Fig.~\ref{fig:fig3}.)}

In Fig.~\ref{fig:fig3}, we increase the magnetic field from left to right. At $B=0$, we have the well-known
spin continuum spectra also with Dirac cones at momenta $\Gamma$, $X$ and
$M$~\cite{maDynamical2018}. Note that without magnetic field, the RPA spin
spectrum has its weight redistributed compared with free spin spectra, due to
the presence of interaction~\cite{maDynamical2018}. With increasing $B$, the
lower part of the spectra appears from the vacuum, making the emergence of the Larmor-Silin spin wave collective mode possible. For both RPA and QMC results at $B=2, 3.4, 4$, we can observe the emergence of quasi-particle like
excitation along side the continuum. Specifically in Fig.~\ref{fig:fig4} (a), each momentum $\mathbf{q}$ of
$S_\pm(\mathbf{q},\omega)$ at $U=2, B=4$ (corresponding to Fig.~\ref{fig:fig3} (h)) is plotted
such that we can clearly see the difference between continuum and collective
mode excitations in QMC+SAC spectra. With larger $B$, there are more momenta
possessing the universal collective mode, with sharp dispersion and finite
life-time.

At small momenta $\mathbf{q} \sim \Gamma$, the collective mode appears at around $\omega \sim B$. And it can be shown that at
exactly the $\Gamma$ point, the RPA treatment with effective $U_\text{eff}$ is
exact, meaning that the collective mode should be exactly at $\omega =
B$~\cite{balentsCollective2020}, independent of Hubbard interaction $U$. This is verified by QMC results in lower panels
of Fig.~\ref{fig:fig3} with different magnetic fields and fixed $U=2$. On the other hand, we also fix magnetic field strength and vary $U$ in QMC. The imaginary-time spin correlation functions at $\mathbf{q}=\Gamma$ for different $U$ with the same
magnetic field $B=4$ are shown in Fig.~\ref{fig:fig4}(b). They all have the same
slope in logscale plot, indicating identical spin excitation gaps whose value
equals to $B$. The mechanism behind it is the Larmor/Kohn
theorem~\cite{oshikawa2002esr}; i.e., when $SU(2)$ rotational symmetry of a spin
system is only broken by
external Zeeman magnetic field, at small momentum there are collective
transverse spin excitations at $B$. This is the Larmor-Silin spin
wave~\cite{silinOscillations1958,silinOscillations1959} discussed in
Refs.~\cite{keselmanDynamical2020,balentsCollective2020,agarwalCollective2023}.
We note that even though the position of the collective mode at the $\Gamma$
point is independent of $U$, this property as well as the emergence of the collective mode
itself are purely interaction effects.

Moreover, we find that at large momentum $\mathbf{q}$, the collective mode resides at the
lower boundary of the continuum, which is not discussed in
Ref.~\cite{agarwalCollective2023}. Compared with the low-energy effective field-theory
calculation, our results of the $\pi$-flux model are based on more realistic band structures,
and our finding of low-energy collective modes at large momenta in the BZ suggests that experiments can look for signatures of such collective modes in a large portion of BZ in realistic materials
such as magnetized graphene and Dirac spin liquid candidates; we note that in inelastic neutron scattering experiments, very often the $\Gamma$ point is blocked due to instrumental difficulties~\cite{zengDirac2023} .
Furthermore, we notice that in RPA near $X(\pi, 0)$, there is a second
collective mode roughly around $B$, while in QMC, a broader peak resides.
The difference may be due to higher order contributions from the interaction
term for larger momenta.

\noindent{\textcolor{blue}{\it Conclusion.}---}
We investigate the spin dynamics of magnetized correlated Dirac fermions
realized in a $\pi$-flux Hubbard model under Zeeman field, with RPA analysis and
unbiased DQMC calculations. We find that when the Dirac fermions are
interacting, there emerges a universal collective transverse spin mode -- the
Larmor-Silin mode -- apart from the usual particle-hole continuum spectra. In
particular, such collective mode not only splits off from the two-spinon
continuum at small momenta as predicted in
Refs.~\cite{keselmanDynamical2020,balentsCollective2020,agarwalCollective2023},
but also appears inside and at the bottom of the continuum at large momenta,
suggesting richer renormalization effects and magnetic field dependence in the lattice model. Our unbiased
results can be used to guide inelastic neutron scattering, electron spin resonance, and other
spectroscopic experiments of quantum Moir\'e materials, topological insulators
and spin liquid materials, under magnetic field.

As pointed out in Ref.~\cite{agarwalCollective2023}, when the interaction is
extended, other collective modes (such as the spin-current collective mode)
would appear. It will be interesting to include the dynamical gauge field
coupling with the Dirac fermion~\cite{xuMonte2019} to find these collective modes of magnetized Dirac spin liquid. We leave it for future work.

{\noindent\it Acknowledgement.--}
We thank Oleg Starykh for the discussion and constructive comments of our manuscript. C.C. and Y. Q. acknowledge the support from National Key R\&D Program of China (Grant No.2022YFA1403400) and from NSFC (Grant No. 12374144).
Y.D.L., C.K.Z., G.P.P. and Z.Y.M. acknowledge the support from the Research Grants Council (RGC) of Hong Kong Special Administrative Region of China (Project Nos. 17301721, AoE/P-701/20, 17309822, HKU C7037- 22GF, 17302223), the ANR/RGC Joint Research Scheme sponsored by RGC
of Hong Kong and French National Research Agency (Project No. A HKU703/22). We thank the Beijng PARATERA Tech CO.,Ltd. (\url{https://cloud.paratera.com}) for providing HPC resources that have contributed to the research results reported within this paper.

% bib part
\bibliographystyle{apsrev4-2}
\bibliography{main}

% SM part
\clearpage
\onecolumngrid

\begin{center}
\textbf{\large Supplemental Material for "Universal collective Larmor-Silin mode emerging in magnetized correlated Dirac fermions"}
\end{center}
%\appendix
\setcounter{equation}{0}
\setcounter{figure}{0}
\setcounter{table}{0}
\setcounter{page}{1}
\makeatletter
\renewcommand{\theequation}{S\arabic{equation}}
\renewcommand{\thefigure}{S\arabic{figure}}
\setcounter{secnumdepth}{3}

In this Supplemental Material, we discuss the implementations of the RPA
calculation of the $\pi$-flux lattice model, and the condition on which the
collective mode appears in the transverse dynamical spin susceptibility, in Sec.
I. Then, we discuss te details of the DQMC simulations, with the HS
decomposition and the SAC from imaginary time dynamic correlation functions to
real frequency spectral functions explained with the necessary references, in
Sec. II. Lastly, we discuss the magnetic spectra at $U>U_c$, in Sec. III.

\section{RPA calculation}
\label{sec:SMI}
The momentum space Hamiltonian of Eq. (1) in the main text without interaction has the
following form
\begin{equation}
\begin{aligned}
H_{\mathbf{k}, \uparrow} &= -\left(\begin{array}{cccc}
\frac{B}{2} + 2\cos \mathbf{k}_y & 1 + e^{-2i\mathbf{k}_x} & \\
1+e^{2i\mathbf{k}_x} & \frac{B}{2} - 2\cos \mathbf{k}_y 
\end{array}\right), \\
H_{\mathbf{k}, \downarrow} &= -\left(\begin{array}{cccc}
-\frac{B}{2} + 2\cos \mathbf{k}_y & 1 + e^{-2i\mathbf{k}_x} & \\
1+e^{2i\mathbf{k}_x} & -\frac{B}{2} - 2\cos \mathbf{k}_y 
\end{array}\right).
\end{aligned}
\label{eq:Hamk}
\end{equation}
The full Hamiltonian matrix $H_\mathbf{k}$ is the direct sum of spin up $H_{\mathbf{k}, \uparrow}$
and spin down part $H_{\mathbf{k}, \downarrow}$, we can further write Hamiltonian
in the diagonal basis
\begin{equation}
\begin{aligned}
H &= \sum_{\mathbf{k},a,b,m,n}c^\dagger_{a,\mathbf{k},\sigma}(U_{\mathbf{k},\sigma})_{am}(D_{\mathbf{k},\sigma})_{mn}(U^{-1}_{\mathbf{k},\sigma})_{nb}c_{b,\mathbf{k},\sigma} \\
&\equiv \sum_{\mathbf{k},\sigma,a}\epsilon_{a,\mathbf{k},\sigma}f^\dagger_{a,\mathbf{k},\sigma}f_{a,\mathbf{k},\sigma}
\end{aligned}
\end{equation}
Because we have two bands for each spin flavor, we use $a,b \in [1,2]$ to denote
band index, and $U$ matrices are $2$ by $2$ such that $m,n \in [1,2]$.
$(D_{\mathbf{k},\sigma})_{mn}=\delta_{mn}\epsilon_{m,\mathbf{k},\sigma}$ with the following
dispersion after considering magnetic field and Hartree shift with effective $U_{\text{eff}}$
\begin{equation}
\begin{aligned}
\epsilon_{a, \mathbf{k}, \uparrow} &= (-1)^a 2t \sqrt{\cos^2 \mathbf{k}_x + \cos^2 \mathbf{k}_y} - \frac{B}{2} - U_{\text{eff}}M,\\
\epsilon_{a, \mathbf{k}, \downarrow} &= (-1)^a 2t \sqrt{\cos^2 \mathbf{k}_x + \cos^2 \mathbf{k}_y} + \frac{B}{2} + U_{\text{eff}}M.
\end{aligned}
\label{eq:dispersion}
\end{equation}
We define $f_{a,\mathbf{k},\sigma}\equiv U^{-1}_{\mathbf{k}}c_{a,\mathbf{k},\sigma}$ with
\begin{equation}
U_{\mathbf{k},\sigma} = \left( \begin{array}{cccc}
-\frac{-2\cos(\mathbf{k}_y)-\sqrt{2}\sqrt{1+\cos^2(\mathbf{k}_y)+\cos(2\mathbf{k}_x)}}{1+\cos(2\mathbf{k}_x)+i\sin(2\mathbf{k}_x)} & -\frac{-2\cos(\mathbf{k}_y)+\sqrt{2}\sqrt{1+\cos^2(\mathbf{k}_y)+\cos(2\mathbf{k}_x)}}{1+\cos(2\mathbf{k}_x)+i\sin(2\mathbf{k}_x)} & \\
1 & 1
\end{array} \right).
\end{equation}
Then the spin raising and lowering operator can be written as
\begin{equation}
\begin{aligned}
S^+(\mathbf{q},a,\tau) &= \sum_\mathbf{k} c^\dagger_{a,\mathbf{k},\uparrow}(\tau)c_{a,\mathbf{k}+\mathbf{q},\downarrow}(\tau) \\
&= \sum_\mathbf{k} \sum_{mn}(U^{-1}_\mathbf{k})_{ma}(U_{\mathbf{k}+\mathbf{q}})_{an}f^\dagger_{m,\mathbf{k},\uparrow}(\tau)f_{n,\mathbf{k}+\mathbf{q},\downarrow}(\tau).
\end{aligned}
\end{equation}
and
\begin{equation}
\begin{aligned}
S^-{\mathbf{q},a,\tau} &= \sum_{\mathbf{k}}c^\dagger_{a,\mathbf{k},\downarrow}c_{a,\mathbf{k}+\mathbf{q},\uparrow} \\
&= \sum_\mathbf{k} \sum_{mn} (U^{-1}_\mathbf{k})_{ma}(U_{\mathbf{k}+\mathbf{q}})_{an}f^\dagger_{m,\mathbf{k},\downarrow}(\tau)f_{n,\mathbf{k}+\mathbf{q},\uparrow}(\tau).
\end{aligned}
\end{equation}
Therefore the bare transverse dynamical spin susceptilibity is
\begin{equation}
\begin{aligned}
\chi^0_{ab}(\mathbf{q},i\omega_n) &= \frac{1}{\beta V} \int^\beta_0 d\tau e^{i\omega_n \tau} \langle  T_\tau S^+_{a,\mathbf{q}}(\tau) S^-_{b,-\mathbf{q}}(0) \rangle \\
&= \frac{1}{\beta V}\sum_{\mathbf{k}\mathbf{k}'} \sum_{mnm'n'}\int^\beta_0(U^{-1}_\mathbf{k})_{ma}(U_{\mathbf{k}+\mathbf{q}})_{an}(U^{-1}_{\mathbf{k}'})_{m'b}(U_{\mathbf{k}'-\mathbf{q}})_{bn'} \langle T_\tau f^\dagger_{m,\mathbf{k},\uparrow}(\tau) f_{n,\mathbf{k}+\mathbf{q},\downarrow}(\tau) f^\dagger_{m',\mathbf{k}',\downarrow}(0) f_{n',\mathbf{k}'-\mathbf{q},\uparrow}(0) \rangle \\
&= \frac{1}{\beta V}\sum_\mathbf{k} \sum_{mn} \int^\beta_0 d\tau e^{i\omega_n \tau} (U^{-1}_\mathbf{k})_{ma}(U_{\mathbf{k}+\mathbf{q}})_{an}(U^{-1}_{\mathbf{k}})_{mb}(U_{\mathbf{k}+\mathbf{q}})_{bn}G^f_{m,\uparrow}(\mathbf{k},-\tau) G^f_{n,\downarrow}(\mathbf{k}+\mathbf{q}, \tau) \\
&= \frac{1}{V}\sum_\mathbf{k} \sum_{mn} \sum_{i\nu_n} (U^{-1}_\mathbf{k})_{ma}(U_{\mathbf{k}+\mathbf{q}})_{an}(U^{-1}_{\mathbf{k}})_{mb}(U_{\mathbf{k}+\mathbf{q}})_{bn} G^f_{m,\uparrow}(\mathbf{k},i\nu_n) G^f_{n,\downarrow}(\mathbf{k}+\mathbf{q}, i\nu_n+i\omega_n) \\
&= \frac{1}{V} \sum_\mathbf{k} \sum_{mn} (U^{-1}_\mathbf{k})_{ma}(U_{\mathbf{k}+\mathbf{q}})_{an}(U^{-1}_{\mathbf{k}})_{mb}(U_{\mathbf{k}+\mathbf{q}})_{bn} \frac{n_F(\epsilon_{m,\mathbf{k},\uparrow}) - n_F(\epsilon_{n,\mathbf{k}+\mathbf{q},\downarrow})}{i\omega_n+\epsilon_{m,\mathbf{k},\uparrow}-\epsilon_{n,\mathbf{k}+\mathbf{q},\downarrow}}.
\end{aligned}
\end{equation}
where we omit $\pm$ in subscript for simplicity and $G^f$ is the $f$-fermion
Green's function.
%After analytical continuation $i\omega_n \rightarrow \omega + i\eta^+$,
%$\chi^0_{ab}$ becomes
%\begin{equation}
%\chi^0_{ab}(\mathbf{q},i\omega_n) =  \frac{1}{V} \sum_\mathbf{k} \sum_{mn} \sum_{i\nu_n} (U^{-1}_\mathbf{k})_{ma}(U_{\mathbf{k}+\mathbf{q}})_{an}(U^{-1}_{\mathbf{k}})_{mb}(U_{\mathbf{k}+\mathbf{q}})_{bn} \frac{n_F(\epsilon_{m,\mathbf{k},\uparrow}) - n_F(\epsilon_{n,\mathbf{k}+\mathbf{q},\downarrow})}{\omega + i\eta^+ +\epsilon_{m,\mathbf{k},\uparrow}-\epsilon_{n,\mathbf{k}+\mathbf{q},\downarrow}}.
%\end{equation}

To get the full spin spectrum, we should keep in mind that the above derivation
is in the folded BZ (the grey BZ in Fig. 1 (b) of the main text). Now we consider the spin correlation on a square lattice (on
site per unit cell), and relate it to above formula,
\begin{equation}
\chi^0(\mathbf{r}_i - \mathbf{r}_j,\tau) = \langle T_\tau S^+(\mathbf{r}_i,\tau)S^-(\mathbf{r}_j,0) \rangle.
\end{equation}
denote $\mathbf{r}_i=\mathbf{r}+\mathbf{v}_a, \mathbf{r}_j=\mathbf{r}'+\mathbf{v}_b$, we have
\begin{equation}
\chi^0(\mathbf{r}+\mathbf{v}_a - \mathbf{r}'-\mathbf{v}_b,\tau) = \langle T_\tau S^+(\mathbf{r},a,\tau)S^-(\mathbf{r}',b,0)\rangle \equiv \chi^0_{ab}(\mathbf{r}-\mathbf{r}',\tau).
\end{equation}
where $\mathbf{v}_a=(0, 0), \mathbf{v}_b=(1, 0)$ are relative position vectors of
sublattice A,B within unit cell.
Perform Fourier transformation, we get the following relation
\begin{equation}
\chi^0(\mathbf{q},i\omega_n) = \sum_{ab}\chi^0_{ab}(\mathbf{q},i\omega_n)e^{-i\mathbf{q}(\mathbf{v}_a-\mathbf{v}_b)}.
\end{equation}

\begin{figure*}[htp!]
\begin{center}
\includegraphics[width=0.8\columnwidth]{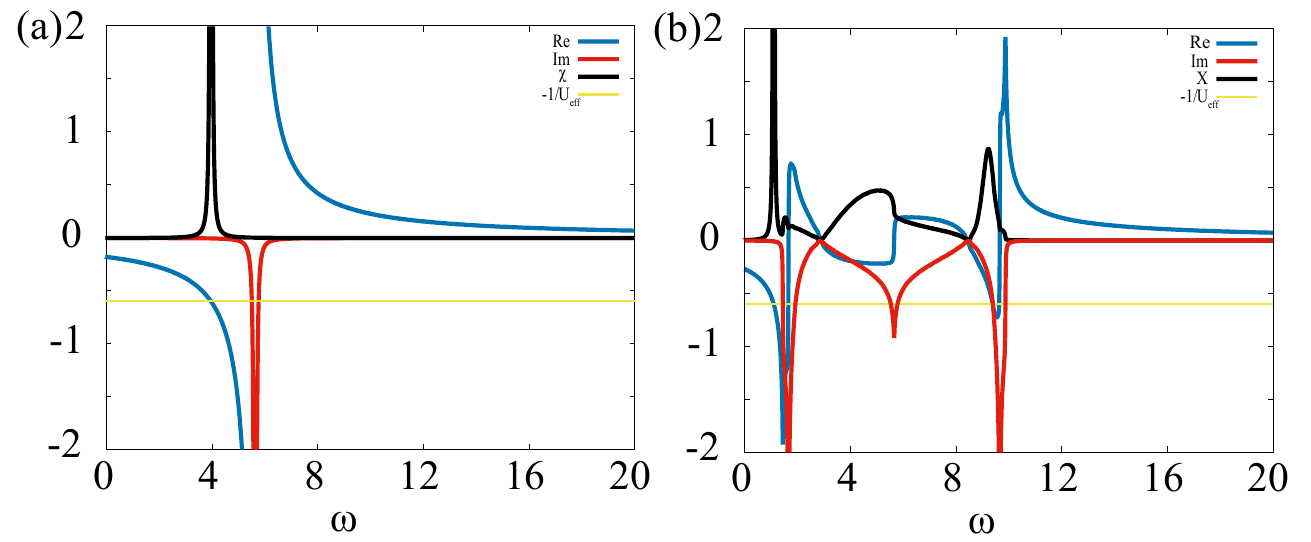}
\caption{Real and imaginary part of bare susceptibility $\chi^{0}_{\pm}$ as well
  as the RPA spectra from Eq.~\eqref{eq:rpa_chi} at the momentum $\Gamma$ (a)
  and $(\pi, \pi/2)$ (b). The data corresponds to $U=2, B=4$ QMC parameter.}
\label{fig:figS1}
\end{center}
\end{figure*}
At last, we can put it into the RPA formula Eq. (3) in the main text, perform
analytical continuation $i\omega_n \rightarrow \omega + i\eta^+$ and
eventually obtain the RPA dynamical spin
susceptibility $\chi_{\pm}(\mathbf{q}, \omega)$. The spin spectra $S_{\pm}(\mathbf{q}, \omega)$
that can be directly compared with QMC+SAC results is
\begin{equation}
  S_{\pm}(\mathbf{q}, \omega) = \frac{-\text{Im}\chi^{0}_{\pm}(\mathbf{q}, \omega)}{[1+U_{\text{eff}}\text{Re}\chi^{0}_{\pm}(\mathbf{q}, \omega)]^2 + [U_{\text{eff}}\text{Im}\chi^{0}_{\pm}(\mathbf{q}, \omega)]^2}.
  \label{eq:rpa_chi}
\end{equation}
by expanding $\chi_{\pm}$ as a function of $\chi^0_{\pm}$.

The origin of the collective Larmor-Silin mode can be elucided from  Eq.~\eqref{eq:rpa_chi}. To have a pole in the denominator, it
requires the bare susceptibility to satisfy $\text{Im}\chi^{0}_{\pm}(\mathbf{q}, \omega)
\rightarrow 0$ and $\text{Re}\chi^{0}_{\pm}(\mathbf{q}, \omega) = \frac{-1}{U_\text{eff}}$
simultaneously. It is thus intuitive to find them near the boundary of spectra
from bare susceptibility $\chi^{0}_{\pm}$. In Fig.~\ref{fig:figS1}, we make
the plot of susceptibility at momentum $\Gamma$ and $(\pi, \pi/2)$ as an example. The
blue line is the real part of $\chi^{0}_{\pm}$, while the red one is the
imaginary part. The black line is the RPA spectra obtained from Eq.~\eqref{eq:rpa_chi}. The thin
yellow line marks the position of $-1/U_\text{eff}$ with $U_\text{eff}=1.67$. At the lower boundary of
the bare spectra ($-\chi^{0}_{\pm}$), the $-1/U_\text{eff}$ thin line crosses the real
part of bare susceptibility, giving RPA spectra (the black line) a sharp peak, this is the collective Larmor-Silin mode at this momentum.

Note that the real part of bare susceptibility is not divergent at the boundary
of bare spectra, it is intuitive to infer that the condition
$\text{Re}\chi^{0}_{\pm}(\mathbf{q}, \omega) = \frac{-1}{U_\text{eff}}$ could not be satisfied for
small enough interaction strength. Therefore, the collective mode might not
emerge as soon as we turn on interaction in the model. One will clearly observe
the collective mode when sufficient interaction and the polarization of the
Dirac dispersions, as shown in the Fig. 3 of the main text. On the other hand, the interaction also shall not be too large, as in that case there would be an
interaction driven quantum phase transition into long-ranged ordered state that
spontaneously breaks spin $U(1)$ symmetry, and the transverse dynamic susceptibility will give rise to the gapless Goldstone mode at the ordered wave vector ($M(\pi,\pi)$ point in our case).

\section{Finite-temperature DQMC method and Simulation details}
\label{sec:SMII}
In this work, we use the finite-temperature determinant quantum Monte Carlo (DQMC) method equipped inverse temperature $\beta=L$ to investigate the ground state properties of a $\pi$-flux Hubbard model with on-site interaction and with external magnetic field. 
In DQMC, we could represent the Hamiltonian -- Eq.(1) in main text -- as $H=H_T+H_U$ with noninteracting part $H_T=H_0+H_B$ and interacting part $H_U$.
The partition function is given as
\begin{equation}
Z=\operatorname{Tr}\left[e^{-\beta H}\right].
\end{equation}
Since $H$ consists of the non-interacting and interacting parts, $H_T$ and
$H_U$, respectively, that are not commute, we should perform Trotter
decomposition to discretize inverse temperature $\beta$ into $L_\tau$ imaginary
time slices ($\beta=L_\tau \Delta\tau$), then we have
\begin{equation}
  Z=\operatorname{Tr}\left[\left(e^{-\Delta_\tau H_U} e^{-\Delta_\tau H_0}\right)^{L_\tau}\right]+\mathcal{O}\left(\Delta_\tau^2\right),
\end{equation}
where the non-interacting and interacting parts of the Hamiltonian are separated.
The Trotter decomposition will give rise to a small systematic error
$\mathcal{O}(\Delta\tau^2)$, we need to set $\Delta\tau$ as a small number to
get accurate results.

$H_U$ contains the quartic fermionic operator that can not be measured directly
in DQMC, to deal with that, one need to employ a SU(2) symmetric
Hubbard-Stratonovich (HS) decomposition, and the auxiliary fields will couple to
the charge density.
In our pare, the HS decomposition is
\begin{equation}
  e^{-\Delta\tau U(n_{i,\uparrow}+n_{i,\downarrow}-1)^{2}}=\frac{1}{4}\sum_{\{s\}}\gamma(s_{i})e^{\alpha\eta(s_{i})\left(n_{i,\uparrow}+n_{i,\downarrow}-1\right)}+\mathcal{O}\left(\Delta_\tau^4\right),
\label{eq:decompo}
\end{equation}
with $\alpha=\sqrt{-\Delta\tau U}$, $\gamma(\pm1)=1+\sqrt{6}/3$,
$\gamma(\pm2)=1-\sqrt{6}/3$, $\eta(\pm1)=\pm\sqrt{2(3-\sqrt{6})}$,
$\eta(\pm2)=\pm\sqrt{2(3+\sqrt{6})}$.

Now, the interacting part is transformed into quadratic term but coupled with an auxiliary field.
Following simulations are based on the single-particle basis $\boldsymbol{c} =
\{c_1, c_2 \cdots c_{N} \}$, so we can use the matrix notation $K$ and $V$ to
represent $H_T$ and $H_U$ operators.
We define the imaginary time propagators
\begin{equation}
\begin{aligned}
  &U_{s}\left(\tau_{2}, \tau_{1}\right)=\prod_{m=m_{1}+1}^{m_{2}} e^{\boldsymbol{c}^{\dagger} V\left(s_{m\Delta\tau}\right) \boldsymbol{c}} e^{-\Delta_{\tau} \boldsymbol{c}^{\dagger} K \boldsymbol{c}}, \\
  &B_{s}\left(\tau_{2}, \tau_{1}\right)=\prod_{m=m_{1}+1}^{m_{2}} e^{V\left(s_{m\Delta\tau}\right)} e^{-\Delta_{\tau} K},
\end{aligned}
\end{equation}
where $m_1 \Delta\tau=\tau_1$ and $m_2 \Delta\tau=\tau_2$.
Then partition function $Z$ can be rewritten as
\begin{equation}
  Z=\sum_{\{s_{\tau} \}} \operatorname{Tr}\left[U_{s}(\beta, 0)\right] \prod_{m=1}^{M} \gamma(s_{m\Delta\tau})e^{-2\alpha\eta(s_{m\Delta\tau})} = \sum_{\{s_{\tau} \}} \operatorname{det}\left[1+B_{s}(\beta, 0)\right] \prod_{m=1}^{M} \gamma(s_{m\Delta\tau})e^{-2\alpha\eta(s_{m\Delta\tau})}.
\end{equation}
Physical observables are measured according to
\begin{equation}
\langle O \rangle = \frac{\operatorname{Tr}\left[\mathrm{e}^{-\beta H} O\right]}{\operatorname{Tr}\left[\mathrm{e}^{-\beta H}\right]}.
\end{equation}
The equal-time single-particle Green function $G_{i,j}(\tau,\tau)$ is given by
\begin{equation}
\left\langle c_{i,\tau} c_{j,\tau}^{\dagger}\right\rangle=\left(1+B_{s}(\tau, 0) B_{s}(\beta, \tau)\right)_{i, j}^{-1}.
\end{equation}
and the dynamical single-particle Green function $G_{i,j}(\tau_1,\tau_2)$ is given by
\begin{equation}
  \left\langle c_{i,\tau_1} c_{j,\tau_2}^{\dagger}\right\rangle=-\left[\left(\mathbf{1}-\mathbf{G}\left(\tau_{1}, \tau_{1}\right)\right) B_{s}^{-1}\left(\tau_{2}, \tau_{1}\right)\right]_{i, j}. \qquad \left( \tau_1 < \tau_2 \right)
\end{equation}
%\pay{Here we need $\tau_2 > \tau_1$ , right?}

Other physical observables can be calculated from single-particle Green function
through Wick's theorem. More technical details of the finite-temperature QMC
algorithms can be found in the
Refs.~\cite{assaadWorld-line2008,xuRevealing2019}. In practice, we set inverse temperature $\beta t=16$ for $L = 16$ lattice and discrete time slice $\Delta\tau=0.1$.

In order to obtain the transverse spin spectra, we firstly measure dynamical
transverse spin correlation function $S_{\pm}(\mathbf{q}, \tau)$
Eq. (4) in the main text and then
make use of SAC to get real frequency information $S_{\pm}(\mathbf{q}, \omega)$.
They are related by
\begin{equation}
S_{\pm}(\mathbf{q}, \tau) = \int^\infty_0 \mathrm{d}\omega\, K(\tau, \omega) S_{\pm}(\mathbf{q}, \omega)
\label{eq:sac}
\end{equation}
where $K(\tau, \omega)$ is the kernel. For boson, the kernel $K(\tau, \omega)$
in Eq.~\eqref{eq:sac} is $\frac{\exp\left(-\tau \omega\right) + \exp\left[ -\left(\beta -
  \tau\right)\omega\right]}{1+\exp\left(-\beta
\omega\right)}$~\cite{sandvik2016constrained}. To get $S_{\pm}(\mathbf{q},
\omega)$ which is in the integrand, SAC technique will be applied.

Let's delve into the intricacies of SAC~\cite{sandvik1998sac,beach2004identifying,sandvik2016constrained,Olav2008,sandvik2019stochastic,shao2023progress,nearlyShao2017}
%~\cite{sandvik1998sac,beach2004identifying,sandvik2016constrained,Olav2008,sandvik2019stochastic,shao2023progress,nearlyShao2017,maDynamical2018,zhou2020amplitude,GYSunDynamical2018,XuZhang2021}
. The concept involves presenting a highly versatile variational ansatz for the spectrum $S_{\pm}(\mathbf{q},
\omega)$ and deriving the associated Green's function $S_{\pm}(\mathbf{q},
\tau)$ in accordance with Eq.~\eqref{eq:sac}. Subsequently, we assess the
agreement between the derived Green's function and the one obtained through QMC, quantified by the parameter $X^2(\mathbf{q})$. The definition of $X^2(\mathbf{q})$ is then elucidated.
	\begin{equation}
	X^{2}(\mathbf{q})=\sum_{i j}\left(\bar{S}_{\pm}(\mathbf{q}, \tau_i) - \int^\infty_0 \mathrm{d}\omega\, K(\tau_i, \omega) S_{\pm}(\mathbf{q}, \omega)\right)\left(C^{-1}\right)_{i j}\left(\bar{S}_{\pm}(\mathbf{q}, \tau_j) - \int^\infty_0 \mathrm{d}\omega\, K(\tau_j, \omega) S_{\pm}(\mathbf{q}, \omega)\right)\end{equation}
	where
	\begin{equation} C_{i j}=\frac{1}{N_{b}\left(N_{b}-1\right)} \sum_{b=1}^{N_b}\left(S^b_{\pm}(\mathbf{q}, \tau_i)-\bar{S}_{\pm}(\mathbf{q}, \tau_i)\right)\left(S^b_{\pm}(\mathbf{q}, \tau_j)-\bar{S}_{\pm}(\mathbf{q}, \tau_j)\right)
	\end{equation}
	and $\bar{S}_{\pm}(\mathbf{q}, \tau_j)$ is the Monte Calro average of Green's functions of $N_b$ bins. $S^b_{\pm}(\mathbf{q}, \tau_i)$ is the Monte Carlo measurement of bin b.
	
	Subsequently, we employ Monte Carlo sampling~\cite{sandvik2016constrained,Olav2008} once again to refine the optimization of the spectral function. We assume a specific form for the spectral function: $S_{\pm}(\mathbf{q},
	\omega)=\sum_{i=1}^{N_{\omega}} A_{i} \delta\left(\omega-\omega_{i}\right)$
	where the weight of the Monte Carlo configuration is given by $
	W \sim \exp \left(-\frac{X^{2}}{2 \,\Theta_T}\right)$.
	Here, $\Theta_T$ serves as an analogue to temperature. Finally, we calculate the average $\langle X^{2}\rangle$ at different $\Theta_T$ through the annealing process. Upon its completion, we can select the converged $\Theta_T$ to fulfill the condition:
	\begin{equation}
	\langle X^{2}\rangle=X_{\min }^{2}+a \sqrt{X_{\min }^{2}}.
	\end{equation}
	Usually we set $a=2$, and the ensemble average of the spectra at such optimized $\Theta_T$ is the final one to present in the main text.

\section{Magnetic spectra at $U>U_c(B)$}
\label{sec:SMIII}
\begin{figure*}[htp!]
\begin{center}
\includegraphics[width=0.8\columnwidth]{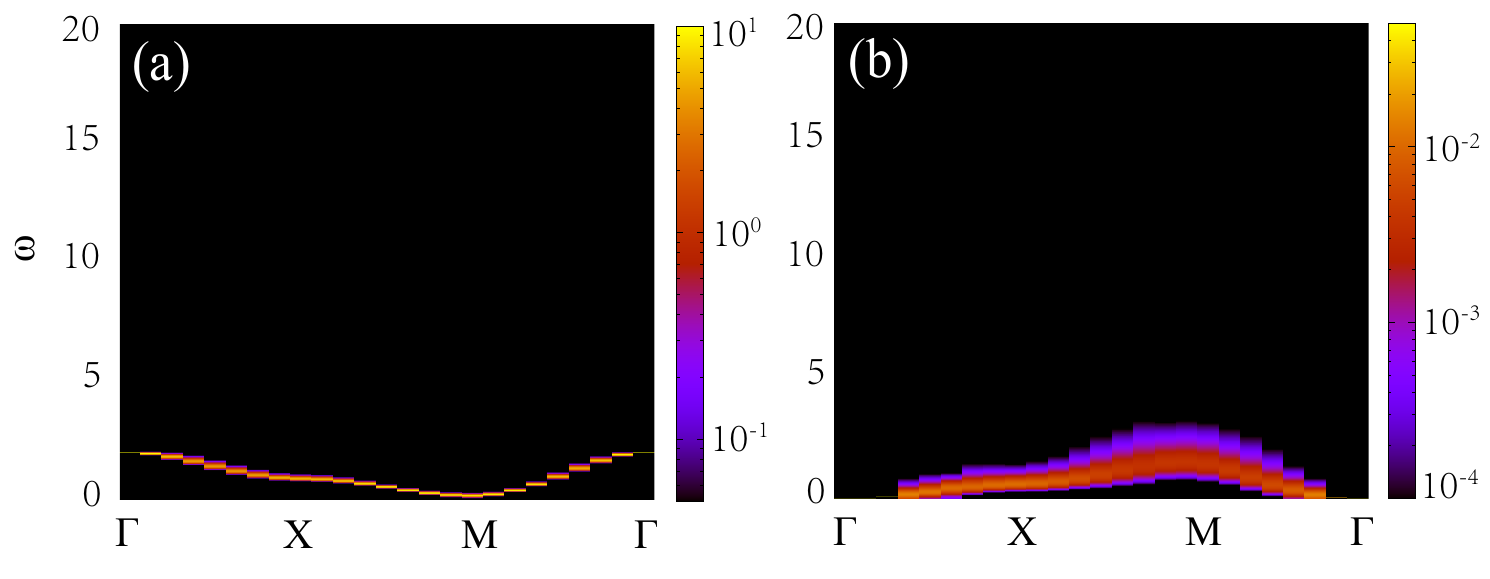}
\caption{Magnetic spectra at $U=8, B=2$ with system size $L=16$ and inverse
  temperature $\beta=16$. (a) transverse channel $S_\pm(\mathbf{q},\omega)$ and
  (b) longitudinal channel $S_{zz}(\mathbf{q},\omega)$. }
\label{fig:figS2}
\end{center}
\end{figure*}

As we increase the on-site Hubbard interaction $U$ of the model with Zeeman
field $B$, a phase
transition occurs from the paramagnetic state to an in-plane
antiferromagnetically ordered state~\cite{bercx2009}.The presence of Zeeman
field $B$, which breaks the $SU(2)$ spin rotational symmetry of fermions,
results in a distinct ordered state at large $U$ that exhibits unique features in the magnetic spectra.
The spectra at large $U$ obtained from DQMC + SAC are shown in
Fig.~\ref{fig:figS2}. The transverse magnetic spectra
$S_\pm(\mathbf{q},\omega)$ resemble spin waves and feature a
gapless Goldstone mode at the ordering momentum $M(\pi,\pi)$.
Meanwhile, $S_{zz}(\mathbf{q},\omega)$ is gapped at $M(\pi,\pi)$.
These spectral results are consistent with previous numerical studies of the AFXY phase~\cite{maDynamical2018}.

\end{document}